Analysis of bimodality in histograms formed from
GALLEX and GNO solar neutrino data

P.A. Sturrock[1]

[1]Center for Space Science and Astrophysics, Stanford University, Stanford, CA 94305-4060
(email: sturrock@stanford.edu)

**Abstract**. A histogram display of the solar neutrino capture-rate measurements made by the GALLEX experiment appears to be bimodal, but that of the follow-on GNO experiment does not. To assess the significance of these results, we introduce a "bimodality index" based on the probability-transform procedure. This confirms that the GALLEX measurements are indeed bimodal (at the 99.98 percent confidence level), and that the GNO measurements are not. Tracking the bimodality index as a function of time shows that the strongest contribution to bimodality comes from runs 42 to 62, i.e. from the time interval 1995.1 to 1996.9. The bimodality index for the first half (runs 1 through 33) is 2.56, whereas that for the second half (runs 33 through 65) is 7.05. Power-spectrum analysis shows a similar distinction: the peaks in the power spectrum formed from the second half are stronger than those in the power spectrum formed from the first half, suggesting that bimodality and rotational modulation are related.

## 1. Introduction

In an earlier article (Sturrock and Scargle 2001), we have drawn attention to the apparent bimodal structure of the histogram formed from capture-rate measurements (that are in turn a measure of the solar neutrino flux) derived from the GALLEX and GNO solar neutrino experiments. The histogram formed from SAGE data also appears to be bimodal, but not as strongly so. The purpose of this article is to investigate GALLEX and GNO data in more detail. The analysis of SAGE data is postponed for a later article.

The key requirement is a procedure for assessing the significance of an apparent bimodality, shown in Section 2, in a capture-rate histogram. Our previous analysis depended upon a binning procedure and involved the consideration of four adjustable parameters, which complicated the significance estimation. In Section 3, we introduce a "bimodality index" that is more robust in the sense that (a) it does not involve binning of the data, (b) it takes into account the asymmetrical error estimates, and (c) it involves no adjustable parameters. We apply this test to GALLEX and GNO data taken separately in Section 4, where we find strong evidence for bimodality in GALLEX data but no such evidence in GNO data.

In Section 5, we introduce a "running bimodality index," that tracks bimodality as a function of time, and use this procedure to analyze GALLEX and GNO data. We find the strongest evidence for bimodality arises from GALLEX runs 44 to 62, i.e. from the time interval 1995.1 to 1996.9.

The significance of bimodality is that one expects sampling from a stationary statistical process to be unimodal. Hence evidence for bimodality may be construed as evidence for variability. A possible source of such variability would be rotational modulation, for which we have found evidence in power-spectrum analyses (Sturrock, Caldwell, and Scargle, 2006). It is therefore interesting to compare the power spectra formed from GALLEX runs 1 to 42 and GALLEX runs 42 to 65. We carry out this comparison in Section 6. Our results are discussed in Section 7.

## 2. GALLEX and GNO Data

We now have available 65 measurements in the GALLEX series (GALLEX I: Anselmann *et al.*, 1993; GALLEX II: Anselmann *et al.*, 1995; Kirsten *et al.*, 1995; GALLEX III: Hampel *et al.*, 1996; GALLEX IV: Hampel *et al.*, 1999) and 58 in the GNO series (Altmann *et al.*, 2000; Kirsten *et al.*, 2003; Altmann *et al.*, 2005). We have taken data from the tables in these articles, except that, for run 42, we adopt for the lower error estimate the value 50 shown in Figure 1 of Hampel *et al.* (1999), rather than the earlier zero value in Table 3 of Hampel *et al.* (1996). Since there are important differences between GALLEX and GNO data [for instance, there is a 5-sigma difference in the error estimates (Sturrock, Caldwell, and Scargle 2006)], we analyze these two datasets separately.

For each run, we are given timing data, a capture-rate estimate for which we now use the symbol $g_n$, and upper and lower error estimates $\sigma_{u,n}$ and $\sigma_{l,n}$, respectively. Hence, for each run, the appropriate form of the pdf for the solar neutrino capture rate $f$ is given by

$$p_n(f \mid g_n)df = \frac{2^{1/2}}{\pi^{1/2}(\sigma_{u,n} + \sigma_{l,n})} \exp\left(-\frac{(f - g_n)^2}{2\sigma_{u,n}^2}\right) df \ \ for \ f > g_n,$$
$$p_n(f \mid g_n)df = \frac{2^{1/2}}{\pi^{1/2}(\sigma_{u,n} + \sigma_{l,n})} \exp\left(-\frac{(f - g_n)^2}{2\sigma_{l,n}^2}\right) df \ \ for \ f < g_n.$$
(1)

We find (Sturrock 2007) that maximum-likelihood analysis yields the following estimates of the solar neutrino capture rate: for GALLEX, 69.8 +/- 7.2 SNU; and for GNO, 54.1 +/- 6.0 SNU.

Histograms formed from GALLEX and GNO capture rate values are shown in Figures 1 and 2, respectively. It certainly appears that the GALLEX histogram is bimodal, but there is no obvious evidence of bimodality in the GNO histogram. The histograms shown in Figure 1 and 2 differ slightly from those shown in Figures 1a and 1b of Sturrock and Scargle (2001). The difference in appearance is due to a difference in binning. This difference indicates that we should attempt to derive a measure of bimodality that is independent of one's choice of bin width and bin location. That is the main aim of this article.

### 3. A Bimodality Index

We now study the distribution of capture-rate values $g_n$ on the assumption that the capture rate is constant. We shall adopt the maximum-likelihood value for the constant value $g_0$, but that is optional. The end result - the significance level of evidence for bimodality - is not much changed if one uses, for instance, the mean value of the capture rate rather than the maximum-likelihood value.

We have shown in a recent article (Sturrock 2007) that if the capture rate is taken to be $g_0$, then the pdf for the measurement $g_n$ is given by

$$p_n(g_n | g_0)dg_n = \frac{2^{1/2}}{\pi^{1/2}(\sigma_{u,n} + \sigma_{l,n})}\exp\left(-\frac{(g_n - g_0)^2}{2\sigma_{l,n}^2}\right)dg_n \text{ for } g_n > g_0,$$

$$p_n(g_n | g_0)dg_n = \frac{2^{1/2}}{\pi^{1/2}(\sigma_{u,n} + \sigma_{l,n})}\exp\left(-\frac{(g_n - g_0)^2}{2\sigma_{u,n}^2}\right)dg_n \text{ for } g_n < g_0.$$

(2)

Note that, by comparison with Equation (1), the two error terms have been interchanged.

If the capture rate were constant, then the pdf for any measurement would be given by

$$P(g)dg = \frac{1}{N}\sum_{1}^{N} P_n(g|g_0)dg. \qquad (3)$$

From this pdf, we may form the cumulative distribution function (cdf)

$$C(g) = \int_{-\infty}^{g} dg' \, P(g'). \qquad (4)$$

We may now use this cdf to carry out a "probability transform" operation

$$\gamma_n = C(g_n). \qquad (5)$$

The important point is that if the $g_n$ values are indeed compatible with the pdf of Equation (2), then the $\gamma_n$ values will be uniformly distributed over the range 0 to 1.

We now wish to examine the possibility that the actual distribution of $\gamma_n$ values is not uniform. We wish to consider, in particular, the possibility that the distribution has two peaks. To this end, we propose using the following "bimodality index"

$$B = \frac{1}{N}\left[\left(\sum_{1}^{N}\cos(2\pi m \gamma_n)\right)^2 + \left(\sum_{1}^{N}\sin(2\pi m \gamma_n)\right)^2\right], \qquad (6)$$

for $m = 2$. This is simply the "power" corresponding to the "frequency" $m = 2$ in a power-spectrum analysis of the $\gamma_n$ values over the range 0 to 1. This is the component in a Fourier analysis that has two peaks and two minima in the range 0 to 1.

If the $\gamma_n$ values are indeed uniformly distributed over the appropriate interval, the index B is distributed exponentially (Scargle 1982),

$$P(B)dB = e^{-B}dB, \qquad (7)$$

so that the probability of obtaining a value B or more by chance is given by

$$P(\geq B) = e^{-B}. \qquad (8)$$

## 4. Bimodality Test of GALLEX and GNO Capture-Rate Measurements

We have applied the probability-transform procedure to the GALLEX and GNO datasets, taken separately. The resulting histograms are shown in Figure 3 and 4, respectively. We see from Figure 3 that the distribution of $\gamma$ values is nowhere near the flat distribution over the range 0 to 1 that we expect. The distribution has two prominent peaks, further indicating that the GALLEX dataset is bimodal. For GNO data, on the other hand, the distribution of $\gamma$ values is fairly flat over the range 0 to 1, so that the GNO dataset may be taken to be unimodal.

We may sharpen these results by computing the bimodality index given by Equation (6). We obtain the value B = 0.54 for GNO data, confirming that there is no evidence of bimodality in the GNO capture-rate measurement, consistent with the appearance of Figures 2 and 4. On the other hand, the bimodality-index has the value B = 8.50 for GALLEX data, confirming that the GALLEX capture-rate measurements are indeed bimodal, as they appear to be in Figures 1 and 3. The probability of obtaining a value of the bimodality index equal to or larger than 8.50 is only 0.0002, indicating that we may be confident, at the 99.98 % level, that the GALLEX capture-rate measurements are indeed bimodal.

## 5. The Running Bimodality Index

It is interesting to track the development of the bimodality index as a function of time. To this end, we replace Equation (6) by

$$B(n) = \frac{1}{n}\left[\left(\sum_{k=1}^{n}\cos(2\pi m\gamma_k)\right)^2 + \left(\sum_{k=1}^{n}\sin(2\pi m\gamma_k)\right)^2\right] \qquad (9)$$

for m = 2. We plot this index as a function of time, for the GALLEX and GNO datasets, in Figures 5 and 6, respectively.

We see from Figure 5 that, for GALLEX data, the running bimodality index increases progressively from run no. 1, with end time 1991.43, up to run no. 44, with end time 1995.11. However, from that time on, the bimodality index increases much more rapidly until run no. 62, with end time 1996.89.

By contrast, we see from Figure 6 that, for GNO data, the bimodality index never exceeds 2.6. Hence, as was evident from Figures 2 and 4, there is no evidence for bimodality of the GNO capture-rate measurements.

## 6. Comparison of Bimodality and the Power Spectrum

In order to compare the bimodality pattern with the power spectrum, we compare results for the first half of the dataset (runs 1 through 33) and the second half (runs 33 through 65). We find that the bimodality index for the first half is 2.56, whereas the bimodality index for the second half is 7.05.

We also show the power spectra formed from the first half and from the second half in Figures 7 and 8, respectively. We see that there appears to be stronger evidence for periodic modulation in the second half than in the first half. Focusing on the rotational band (12.5 to 13.8 $yr^{-1}$; see for instance Sturrock, Caldwell, and Scargle 2006) we find a peak in the second half at 13.13 $yr^{-1}$, with power S = 7.13. By contrast, in the power spectrum formed from the first half, the strongest peak in that band is found at 13.54 $yr^{-1}$, with power S = 3.35. However, there are other differences in the power spectra – for instance, the mean power over the band 0 to 30 $yr^{-1}$ is 0.92 for the first half of the GALLEX dataset, but 1.61 for the second half. We defer a more detailed investigation of the evolution of the power spectrum as a function of time for a later article.

### 7. Discussion

It is interesting to compute expressions analogous to Equation (6) for other values of m. For m = 1, we obtain B = 0.55 for GALLEX data, and B = 2.75 for GNO data. For m = 3, we obtain B = 0.36 for GALLEX data, and B = 1.06 for GNO data. These results give no evidence of "unimodality" (a more marked "clumping" of count-rate values than one would expect from the error estimates) or "trimodality."

The results of Section 6 suggest that there may be an association between bimodality in the capture-rate histogram and rotational modulation. Such a possible association was the stimulus for our early study of bimodality (Sturrock and Scargle, 2001). However, it is pertinent to note that runs 1 through 39 are derived from GALLEX I and GALLEX II, whereas runs 40 through 65 are derived from GALLEX III and GALLEX IV. Hence the differences in the bimodality and power-spectrum results could be due in part to possible systematic differences between the first two experiments and the second two experiments. It does not seem possible to assess this issue from the available published data.

We have previously noted (Sturrock, Caldwell, and Scargle (2006); see Figure 3) that there appears also to be a large secular variation in the count rate measured by the GALLEX experiment, from a low of about 40 SNU to a high of about 100 SNU. Hence there are at present three lines of evidence pointing towards variability of the solar neutrino flux as measured by the GALLEX experiment: apparent bimodality of the count rate, apparent rotational modulation, and apparent secular variation.

If we accept the evidence that the GALLEX capture-rate histogram is indeed bimodal, it is interesting to try to estimate the locations of the two peaks. We have examined separately 29 runs with capture-rate values between 20 and 80 SNU, and 23 runs with capture-rate values between 80 and 140 SNU. A maximum-likelihood analysis of these two datasets yields the estimate $51.5 \pm 9.2$ *SNU* for the lower peak and $111.2 \pm 13.9$ *SNU* for the upper peak. If we accept these as valid estimates of two states of the solar-neutrino-measurement process, we see that they differ by 59.7 SNU, and that the expected error of the difference is 16.7 SNU. The ratio yields a 3.6 sigma difference in the two peaks.

According to Pulido (2007), a count rate of 110 SNU would be incompatible with the LMA version of the MSW process (Vignaud, 1992). We hope to obtain a more definitive estimate of the maximum value of the count rate from a future analysis.

## Acknowledgements

We wish to thank Luciano Pandola and Joao Pulido for helpful comments on this work, which was supported by NSF Grant AST-0607572.

FIGURES

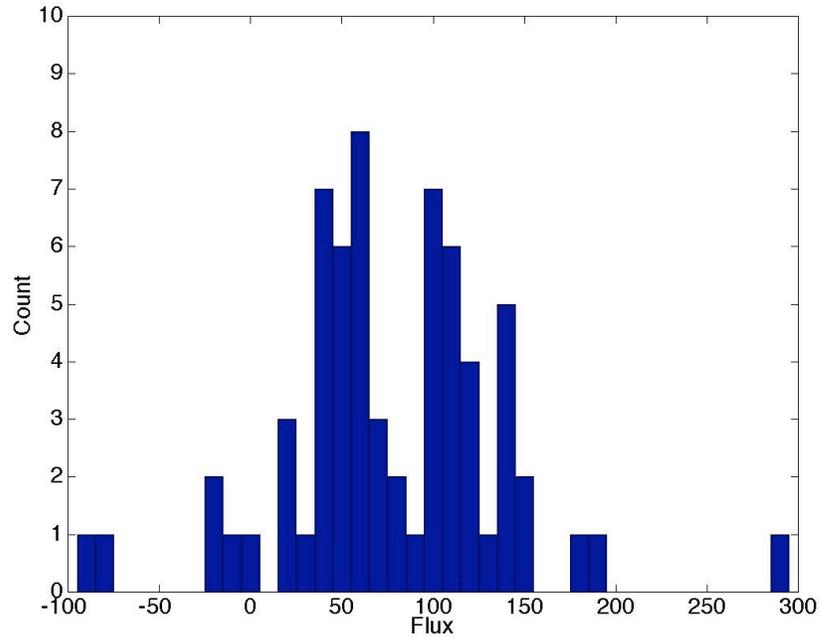

Figure 1. Histogram formed from 65 GALLEX capture-rate measurements.

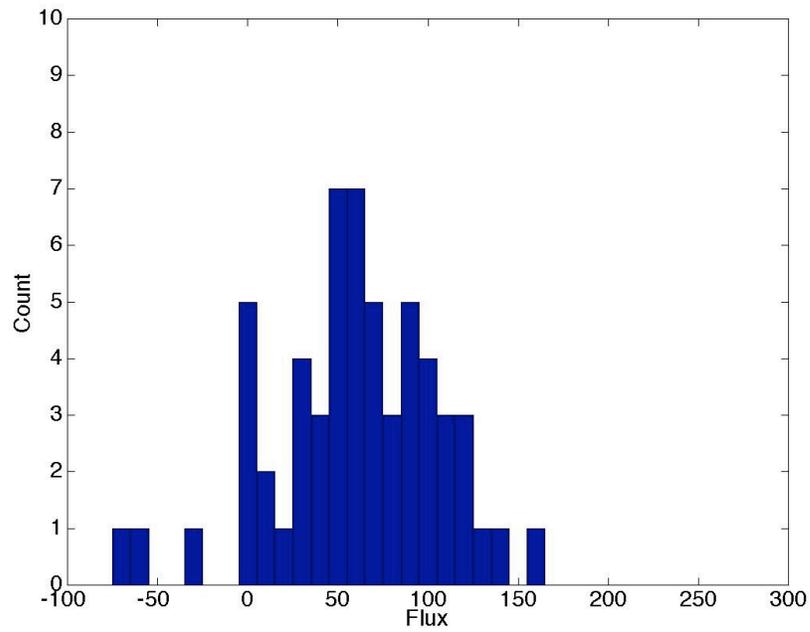

Figure 2. Histogram formed from 58 GNO capture-rate measurements.

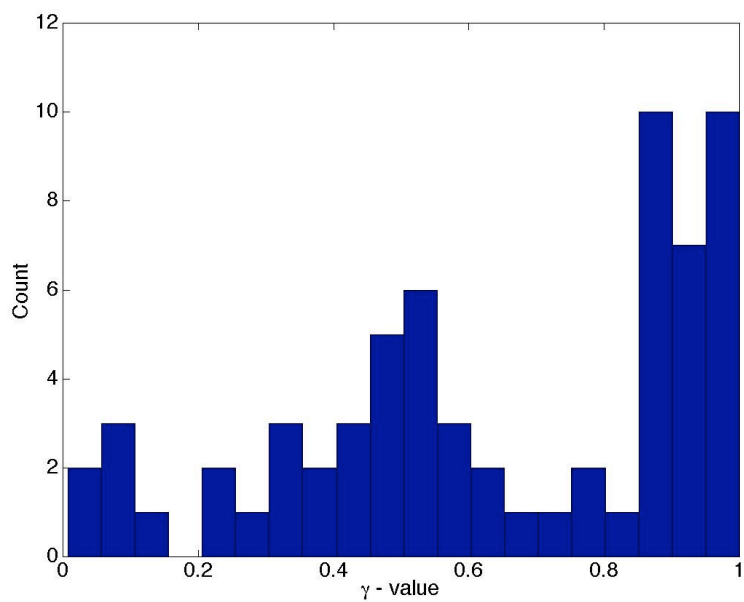

Figure 3. Histogram formed from probability-transform values derived from 65 GALLEX capture-rate measurements.

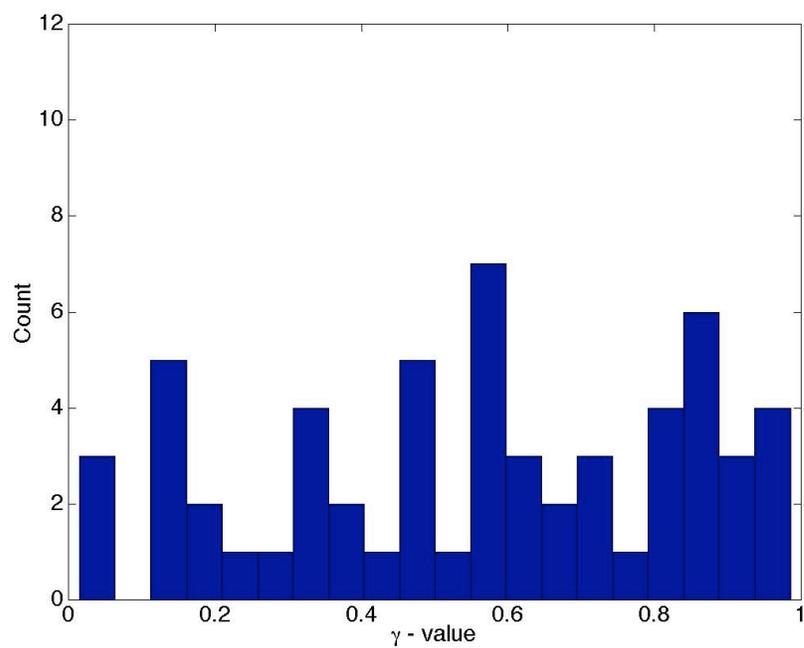

Figure 4. Histogram formed from probability-transform values derived from 58 GNO capture-rate measurements.

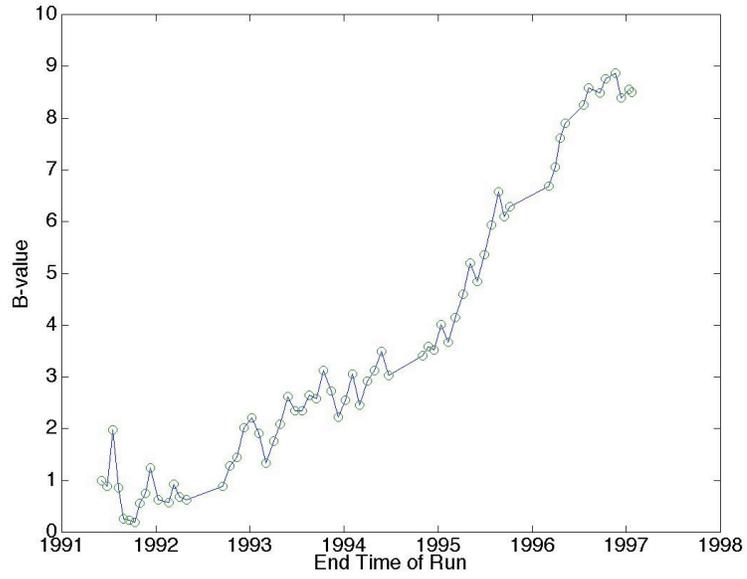

Figure 5. The running bimodality index formed from GALLEX capture-rate measurements.

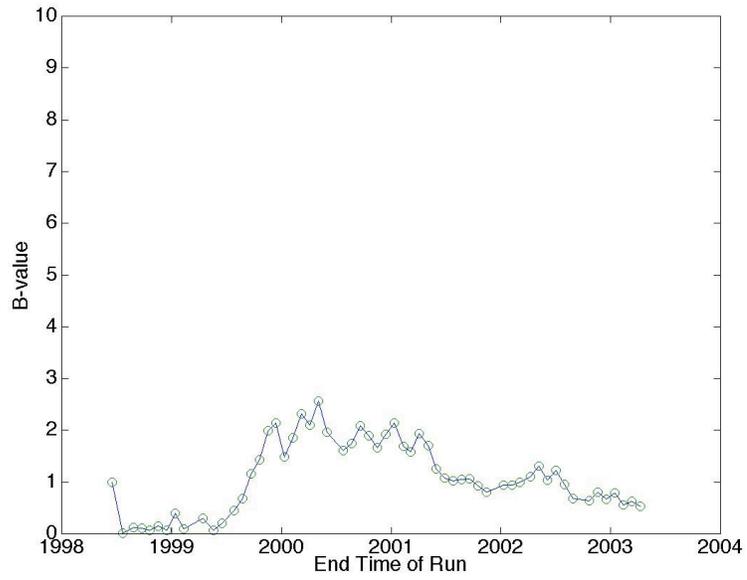

Figure 6. The running bimodality index formed from GNO capture-rate measurements.

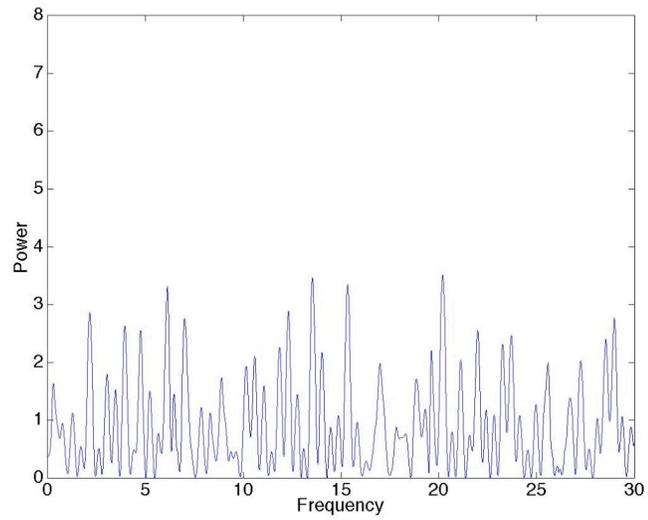

Figure 7. The power spectrum formed from the first half of the GALLEX dataset.

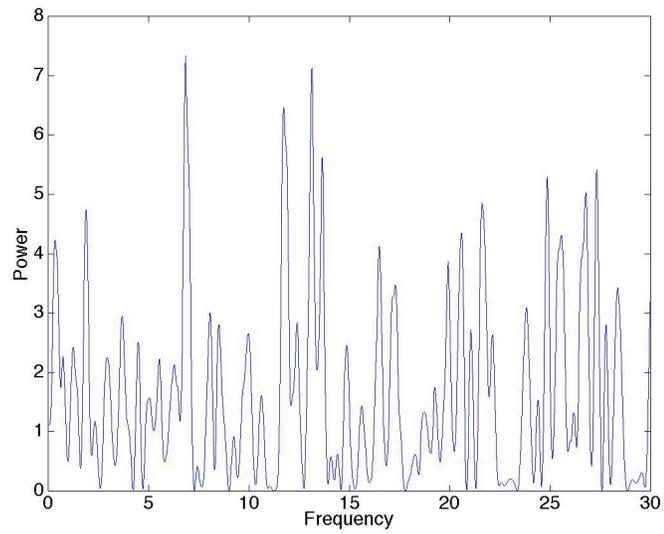

Figure 8. The power spectrum formed from the second half of the GALLEX dataset.